\documentclass{aa}
\usepackage{times}
\usepackage{graphics}
\usepackage{psfig}
\begin{document}
\newcommand {\fmag}   {\mbox{$\,^m\!\!\!\!.\ \,$}}
\newcommand {\mg}     {\mbox{$\mathrm{Mg}_2\,$}}
\newcommand {\lrho}     {\mbox{$\log(\rho)\,$}}
\newcommand {\sig}     {\mbox{$\sigma_0\,$}}
\newcommand {\kms}    {\mbox{km sec$^{-1}$}}
\newcommand {\rotsup} {\mbox{$V_{\rm max}/\sigma_0\,$}}
   \thesaurus{11     
              (11.05.1;  
               11.06.1;  
               11.06.2;  
               11.07.1;  
               11.19.5)} 
   \title{Effect of the large scale environment on the stellar content 
          of early-type galaxies
         \thanks{Based on observations collected at the 
          Observatoire de Haute-Pro\-vence}
         }

   \author{
           Ph.~Prugniel
           \inst{1},
           V.~Golev
           \inst{1,2}
           \and
G. Maubon \inst{1}\fnmsep
          }

   \offprints{Ph.~Prugniel (prugniel@galaxies.univ-lyon1.fr}

   \institute{CRAL-Observatoire de Lyon (CNRS: UMR 142), 
              F-69561 St-Genis-Laval Cedex, France         
          \and
              Department of Astronomy and Astronomical Observatory,
              University of Sofia, P.O.Box 36,
              BG-1504 Sofia, Bulgaria
             }

   \date{Received March 25, 1999; accepted May 25, 1999}

   \maketitle

   \begin{abstract}
In order to study the influence of the environment on the stellar
population of early-type galaxies we have analyzed the relations between
the density of the environment \lrho and
the residuals to the \mg -- \sig relation, $R_a$, and between 
\lrho and the residuals to the Fundamental Plane, $R_f$.
Our sample of galaxies covers the range of densities, between field
galaxies and poor clusters. 
We confirm the existence of a slight environmental bias on the mean stellar
population and we show that this effect is due to a small fraction of galaxies
containing a younger stellar sub-population while the majority of galaxies,
in any environment, are equally old. 
The galaxies which are likely to contain a young
stellar sub-population (negative $R_a$ and $R_f$)
are preferentially found in the less dense environment.

This segregation of the stellar population may reflect the morphology-density
relation or may result from a higher rate of recent star formation activity 
in sparse environments. Using the rotational support, \rotsup, to assess the
presence of a disk and to parameterize the morphology, we subtracted
the contribution of the morphological segregation to population segregation.
It is not yet possible to rule out that the morphology-density relation is 
at the origin of the population-density relation.
      \keywords{galaxies: general --
                galaxies: elliptical and lenticular, cD --
                galaxies: fundamental parameters --
                galaxies: stellar content -- 
                galaxies: formation 
               }
   \end{abstract}

%

\section{Introduction}\label{sect_intro}

The comparison of the stellar populations at different distances 
(e.g., Bender et al. \cite{Be+93}, Ziegler \& Bender \cite{ZB97},
Stanford et al. \cite{Stan98}) 
shows that early-type galaxies (E and S0) are essentially old and 
undergo only a passive evolution. Even in sparse environments, where
hierarchical models of galaxy formation predict a younger age, the
mean ages are not dramatically younger (J{\o}rgensen \cite{Jor97},
Bernardi et al. \cite{Ber98}).
However, a fraction of elliptical galaxies contains also a younger
component. This is apparent from statistical analyses, eg. 
Forbes et al. (\cite{For98}), and strong evidence comes from the observation of
merger remnants (e.g., NGC 7252, Schweizer \cite{Sch82}) which will turn
into ``normal'' ellipticals after a couple of gigayears.

These intermediate-age sub-populations 
provide the framework for understanding the connection 
between {\it fine structures}  and spectral peculiarities 
(Schweizer et al. \cite{S+90}) and probably also for interpreting
the {\it peculiar velocities} found in clusters (Gregg \cite{Gr92}) 
and residuals to the Fundamental Plane (Prugniel \& Simien \cite{PS96}). 

It is generally accepted that the rate of star formation in 
early-type galaxies is enhanced
in low-density environment (Schweizer \& Seitzer \cite{SS92},
de Carvalho \& Djorgovski \cite{dCD92},
Guzm\'an et al. \cite{G+92},
Rose et al. \cite{Ros94},
J{\o}rgensen \& J{\o}nch-S{\o}rensen \cite{JJ98}). It results in a
population -- density relation: The metallic features in the spectra (eg. \mg) 
are weaker and the Balmer lines stronger in low-density regions.

However, the interpretation of this environmental dependence is not 
straightforward. 
For example, Mehlert et al. (\cite{M+98}) interpret the dependence of \mg\
on the distance to the center of the Coma cluster, previously 
found by Guzm\'an et al. (\cite{G+92}), as a bias in the morphological
classification. The ``young'' early-type galaxies 
found in the outskirts of Coma
would actually be lenticular galaxies, and the population segregation 
would simply reflect the morphological segregation.

Because the stellar population varies with the Hubble type,
at least part of the population -- density relation is likely a
by-product of the morphology-density relation 
(Dressler \cite{Dr80}, \cite{Dr84}, Whitmore et al. \cite{WGJ93}; 
the fraction of early-type galaxies is higher in dense regions).
Conversely, the population segregation may bias the morphological 
classification (for spiral galaxies) and hence contribute to
the morphological segregation (Koopmann \& Kenney \cite{KK98}).

The question left open is to determine the fraction of the population
segregation due the morphological segregation.
It is connected with that of the origin of the morphological
segregation (Martel et al. \cite{Mar98}). This segregation may be due to 
initial conditions, morphological evolution (due to gas stripping and
mergers), or a combination of both. The first hypothesis is difficult to 
defend in the light of the existence of clear cases of merger remnants.
If the environment is responsible
for a significant morphological evolution, it will also be at the origin of
a population segregation (at a given morphological type)
because the morphological transformations are
accompanied with star formation.

In elliptical galaxies, the ``young'' stellar sub-populations are likely due 
to merging with a gas-rich companion that occurred at maximum a 
couple of gigayears ago. In lenticular galaxies, the recent populations may be
the result of the residual star formation in the gaseous disk. The
first class of object will not be subject to the morphological segregation,
while the second will be (in this case the morphology is related to the
importance of the disk). Subtracting the contribution of the morphological
segregation to the observed population segregation, would in principle allow
to determine the present rate of environmentally triggered star formation.

The aim of this Letter is to study the population segregation in the
sample of nearby early-type galaxies presented in Prugniel \& Simien
(\cite{PS96}), and to address the question of its origin.
We will first parameterize the density of the environment
using the HYPERCAT database
(http://www-obs.univ-lyon1.fr/hypercat). 
Then we will use two diagnostics to detect the
presence of a young stellar sub-population. 
The first one is the analysis of the 
residuals, $R_a$, to the well-known relation between the
magnesium line strength index \mg and the central velocity dispersion \sig
(Terlevich et al. \cite{Ter81}, Bender et al. \cite{Be+93}).
The second approach similarly analyses  
the residuals to the Fundamental Plane (FP), $R_f$, following the line of 
Prugniel \& Simien (\cite{PS94}, \cite{PS96}, and \cite{PS97}, 
collectively referred as PS). 
The analysis of the rotational support of the galaxies where a 
young sub-population is detected will allow to determine if the age
segregation is linked to the morphology-density relation. 

%
\section{Analysis}

Our sample of galaxies, described in PS, consists in
nearby early-type galaxies in different environments. Photometric
and kinematic observations were obtained using the
1.20m and 1.93m telescopes of Observatoire de Haute Provence and the 
{\em CARELEC} long-slit spectrograph.
This material is already presented in details in
Prugniel \& H\'eraudeau (\cite{PH98}), Prugniel \& Simien (\cite{PS94}), 
Simien \& Prugniel (\cite{SP97a}, \cite{SP97b}, and \cite{SP97c}), and 
Golev et al. (\cite{Go+99}). 
All the data are also available through HYPERCAT. Both \mg (Lick system)
and \sig are aperture corrected and standardized to an homogeneous system
(see Golev \& Prugniel \cite{GP98}).

The sample includes bona-fide elliptical and lenticular galaxies 
as well as merger remnants. The classification 
was made on the basis of literature assessments and presence of peculiarities:
galaxies with morphological disturbances, dust-lanes, post-mergers ... were
rejected from the bona-fide elliptical subsample 
(Prugniel \& Simien \cite{PS96}).
This classification is clearly subjective, and some of 
the bona-fide ellipticals may actually
hide a disk and should be re-classified as lenticular. This is a general
drawback of the morphological classification of galaxies (see Kormendy \&
Bender \cite{KoBe96}, Andreon \cite{AND98}). In this paper, 
the bona-fide ellipticals subsample will be used as a reference for 
determining the FP and \mg -- \sig relations.

\subsection{Parmeterizing the density}\label{subsect_par}

To study environmental effects, we will define a parameter to
measure the density of the environment. Ideally, the density of the environment
should be the number of galaxies per megaparsec cubes locally measured
in a given volume around a galaxy. 
For our purpose, the smoothing-volume will be the
group or cluster the galaxy belongs to. The underlying idea is 
to characterize the mean environment of a galaxy over its $10^{10}$
years life in order to connect the density of this environment with
the stellar population built over the same period.

We will start from the sample
of galaxies with measured radial velocity smaller than 9000 \kms extracted
from HYPERCAT (the velocity data compilation primarily comes from the 
LEDA database\footnote{
http://www-obs.univ-lyon1.fr/leda
   }). It contains 22689 galaxies, but is
not complete in apparent magnitude or diameter. Extracting a 
magnitude-limited whole-sky sample would produce a list of galaxies too
restricted to allow proper determinations of the density of the environment. 
We prefer to apply
completeness corrections to the present sample. Since the number of galaxies
per 1000 \kms shell is almost constant in the sample, assuming a uniform
distribution of the galaxies in all the volume suggests a completeness 
correction: $\rho_{\rm c} = \rho_{\rm u} * (V_{\rm dist}/1000)^2$ 
(where $\rho_{\rm u}$ and $\rho_{\rm c}$ are 
respectively the un-corrected and corrected densities, in relative units, 
and $V_{\rm dist}$ 
the distance of the group or cluster in \kms). We do not apply correction for
galaxies nearer than $V_{\rm dist}=1000$ \kms.
This correction supposes that the list of galaxies with known redshift
uniformly samples the real distribution of galaxies. This is a crude
hypothesis, since most redshift surveys concentrated on limited regions
(in particular clusters), and this could lead to an overestimation of
the density in the regions of deep redshift surveys. To check 
the magnitude of this bias, we compared the redshift sample with the UGC and
ESO samples (which are diameter limited). We found, that the density 
projected on the sky of the redshift catalogue, normalized
to the density in the UGC and ESO samples, fluctuates by a factor 2 rms
(the density was smoothed in 5 degree diameter disks). We did not attempt
to use this result to modify the completeness correction, because that would
have supposed to apply a further correction to the UGC/ESO samples, making
assumptions on the luminosity function of galaxies, which would carry 
uncertainties not smaller than a factor 2 also.

The algorithm to group the galaxies is described in 
Golev \& Prugniel (\cite{GP98}). It associates a group to all
the galaxies in the sample, returning the mean (flow) velocity, the radius
and the number of galaxies grouped. This is used to compute the density
and also the aperture corrections which are applied to the \mg
and \sig data.
Rescaled to the mean density of Virgo, 
this algorithm gives $\rho = \rho_{\rm c}/\rho_{\rm c}\mathrm{(Virgo)} = 0.6$ for Fornax, 
4 for Coma and about 0.3 for Leo or NGC 5846 groups.
These values restore the hierarchy of concentrations between these different
groups and clusters, thus validating our measurements of $\rho$. 
The density provides a smooth parameterization of the field-group-cluster
classification.
Our sample covers mostly the range of low densities, from field galaxies to
poor clusters.

\subsection{Residuals to the \mg - \sig relation}

The existence of a tight correlation between \mg and \sig was discovered
by Terlevich et al. (\cite{Ter81}) and further discussed in Burstein
et al. (\cite{Bur88}) and Bender et al. (\cite{Be+93}). 
The slope of this relation is clearly due to the metallicity, and not age,
as assessed by the constancy of the slope of the color-magnitude
relations out to z$\approx$0.9 (Stanford et al. \cite{Stan98},
Kodama et al. \cite{Kod98}). 
At variance, the spread
around this relation is likely due to the contribution of young
sub-populations which results in a skewness of the residuals 
(Burstein et al. \cite{Bur88}, Prugniel \& Simien \cite{PS96}).

We have used the subsample of bona-fide ellipticals to fit
the relation between \mg and \sig taking properly into account the
errors on both coordinates. The residuals, $R_{a}$, fitted on 308 objects,
are defined as:
  \begin{equation}\label{eq1}
    R_a = \mg - 0.225 \pm0.052 \times \log (\sigma_{0}) + 0.235 \pm0.083
  \end{equation}
 
\noindent 
The estimated error on calculated \mg is $0.026$.
This fit is in agreement with Guzm\'an \& Lucey (\cite{rguz193}) and
with Davies et al. (\cite{rldav187}) who both find a slope of 
$0.20 \pm 0.05$.

Separating this sample in two parts according to the density, we find
$<R_{a}> = 0.003 $ (154 galaxies) and $<R_{a}> = -0.004 $ (154 galaxies)
respectively for the high- and low- density subsamples.
The difference, $0.007 \pm 0.005$, perfectly agrees with similar estimates
by J{\o}rgensen (\cite{Jor97}) and  Bernardi et al. (\cite{Ber98}).

\begin{figure}
      \resizebox{\hsize}{!}{\includegraphics{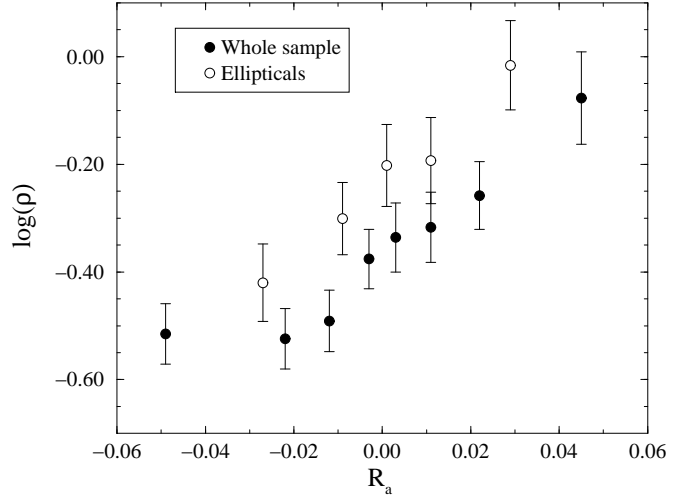}}
      \caption[]{Mean density of the environment \lrho as a function of 
the residuals $R_a$ to the \mg -- \sig relation, $R_a$. 
The filled symbols represent the whole sample, 
while the open symbols are the elliptical galaxies alone. Each point represents 
average \lrho for 80 (whole sample) or 40 (elliptical) galaxies. The error 
bars are the 1-$\sigma$ uncertainty on the position of each point.
                 }
         \label{fig1}
   \end{figure}

In the present analysis, we are interested in the nature and environment
of early-type galaxies hosting a young sub-population. For this purpose,
Fig.~\ref{fig1} presents the average \lrho as a function
of $R_{a}$. The galaxies are grouped in bin of similar $R_{a}$
(ie. presumably similar stellar population).
The trend for galaxies with most negative residuals to be located 
in lower density environment is clear.

We have over-plotted in Fig.~\ref{fig1} the subset of elliptical galaxies, it is
interesting to note that the same tendency is observed, but with a slight
offset toward higher density at a given $R_{a}$. This offset probably 
reflects the morphology-density relation:
For a given \mg, elliptical galaxies reside in denser environments than
lenticular galaxies.
In addition, assuming that our sample of elliptical galaxies is
morphologically homogeneous, ie. not contaminated by S0 galaxies,
the fact that they are also segregated indicates that the population-density
relation may be more than a by-product of the morphology-density relation.

Negative $R_{a}$ values may be due either to a low \mg or symmetrically
to a high \sig. While the first hypothesis is the most favored in the
frame of the usual interpretation, the second may result from the
dynamical evolution of galaxies harassed by gravitational encounters.
In order to check this possibility we will analyze the residuals to the 
FP: A low \mg or a high \sig will result in opposite residuals.

%

\subsection{Residuals to the Fundamental Plane}\label{sect_fp}

%

We performed a fit of the FP using the subsample of 291 {\em bona-fide} 
elliptical galaxies (17 galaxies were excluded from the 308 galaxies sample 
because of the lack of reliable photometry):

  \begin{equation}\label{eq2}
    R_f = 2\/\log\sigma_0 + 0.2 (1 + 2\/\beta) M_B +0.2\mu_e + \eta\/. 
  \end{equation}
 
\noindent $R_f$ is the residual to the FP, \sig, the central velocity dispersion
(in \kms) $M_B$ and $\mu_e$ respectively the asymptotic magnitude and mean
surface brightness within the effective aperture in the B band, $\beta$ and
$\eta$ are the free parameters. Their best fit values are:
$\beta = 0.20\pm0.02$, $\eta = -3.2\pm0.2$, 
similar to the values previously reported in PS. 

Note that we fitted the classical FP, ie., we did not include
the additional terms accounting for the stellar population, the rotational
support and the non-homology of the spatial structure (see PS). 
Since we precisely want to
use the residuals to the fundamental plane as a parameterisation of the stellar
population, we did not include the first scaling relation in the FP equation.
Including the two latter would have dramatically reduced the size 
of the sample and is not useful to the present goal. 
%
   \begin{figure}
      \resizebox{\hsize}{!}{\includegraphics{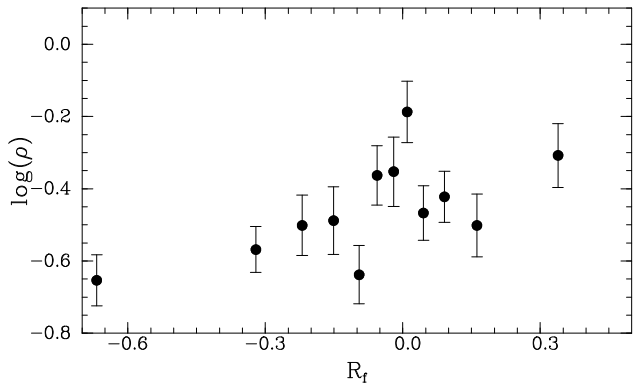}}
      \caption[]{The mean relative density of the environment
\lrho plotted as a 
                 function of the residuals $R_f$ to the Fundamental Plane. 
Each point corresponds to a bin of 40 galaxies, the error bars are the
$1-\sigma$ uncertainty on the position of the points.
                 }
         \label{fig2}
   \end{figure}
%

In Fig.~\ref{fig2} we present the average of \lrho
as a function of $R_f$.
Each point on the figure represents 40 galaxies. 
There is a (somewhat marginal) tendency for the galaxies with the more 
negative residuals to be found in lower density environments.

Splitting the sample in two parts, we find for
$R_f<-0.1$, \lrho$=-0.35$ (154 galaxies), and for $R_f<-0.1$, 
\lrho$=-0.19$ (264 galaxies). The uncertainty on the mean \lrho 
being about 0.06, the difference is significant at the 3 $\sigma$ level.

Repeating this on arbitrary extracted subsamples of half the total size
produce the same result. Therefore, we conclude that the trend is significant.
Since there is no reason why the measurement errors could be connected 
with the environment, we conclude that the relation is physically significant.

Combining the results from the $R_a$ and $R_f$ analyses, we confirm the
existence of a population segregation, and extend it to the low-density
environments.

\section{The origin of the population segregation}

Our result fits in the frame of the current paradigm suggesting a reduced
rate of star formation in high density environments. The morphology-density
relation, the HI deficiency of cluster spirals or the relative
isolation of shell galaxies are classically interpreted as the result
of past gravitational interactions which both stimulated the star formation
earlier in the life of the galaxies and stripped the dense gas, 
prohibiting present-epoch star formation.

We will now try to extract the fraction of the population segregation
due to recent merging or accretion of a gas-rich companion, which 
cannot be accounted for by the morphological segregation.

\subsection{Evidence for recent merging events and star formation triggered by 
gravitational encounters}

Apart the violent cases of star forming galaxies often associated
with strong interactions or mergers, the ``weak'' interactions
have also been related with the presence of a young stellar component.
In particular, Schweizer et al. (\cite{S+90}) and Schweizer \& Seitzer 
(\cite{SS92}) found clear correlations between the anomalies on the colors,
\mg, $H\beta$ and CN, and the fine structure index $\Sigma$ 
(indicating the amount of shells, ripples, boxiness, etc.). 
Gregg (\cite{Gr92}) showed that the $\Sigma$-index is
also correlated with the residuals to the FP 
({\em peculiar velocities} in his terminology). To summarize, the
presence of a ``young'' sub-population is clearly
associated with merging events or gravitational encounters. Hence,
this population effect is expected to depend on the environment.

Indeed, Guzm\'an et al. (\cite{G+92}) found a difference 
between the galaxies located in the outer and inner regions of the Coma 
cluster.
The former have negative residuals (i.e. they are younger or less metallic).
Bower et al. (\cite{Bo+92}) found a difference between the Coma and Virgo 
clusters, but this could also be an observational effect due to the difference 
in the projected slit sizes. 
Recently J{\o}rgensen \& J{\o}nch-S{\o}rensen (\cite{JJ98}) showed that 
the colors and the absorption-line indices of E and S0 galaxies belonging 
to the poor cluster S\,639 indicate that the stellar populations in these 
galaxies are probably younger (or less metallic) than those in rich clusters.
This is in agreement with the results by de 
Carvalho and Djorgovski (\cite{dCD92}), J{\o}rgensen (\cite{Jor97}) and
Bernardi et al. (\cite{Ber98}).

It is important to note that these previous works searched for mean-age 
differences as a function of the environment. For instance, 
Bernardi et al. (\cite{Ber98}) found a zero-point difference on the
\mg -- \sig relation of $\delta(\mg) = 0.007 $ between cluster and field
ellipticals, compatible with our comparison between the high- and low-density
subsamples. Assuming a single burst stellar population model 
(Worthey \cite{Wor94}), this corresponds to an age difference of about 1 Gyr.

In the present work, we found a much stronger difference
when grouping the galaxies according the $R_a$ or $R_f$ residuals.
The galaxies containing a younger stellar population are clearly found in
lowest-density environment, but, in any environment,
the majority of the early-type galaxies have only an old population. 
The rare galaxies containing young stars form the tail of negative residuals
noted in Burstein (\cite{Bur88}) and Prugniel\& Simien (\cite{PS96}).
This same skewness is also apparent in the distribution of the color-magnitude
diagrams. It is also reminiscent of the population of blue galaxies in clusters
(Butcher-Oemler effect; Arag\'on-Salamanca et al. \cite{Ara91}).
If we compare the median of $R_a$, less sensitive to the skewness, 
instead of the average, the difference between the 
high- and low-density subsamples vanishs.

We also tried to connect the residuals with the small-scale density of 
the environment
by weighting the galaxies according to their separation on the sky 
when computing
the density, with the scope of defining a ``strength of the tidal field''. 
We experimented different weightings, but failed to find
evidence for an effect related with the ``tidal'' field. 
Only the large scale environment correlates with the stellar 
content.

The population segregation
may still be contaminated by the morphological segregation.
For example,
Kuntschner \& Davies (\cite{KD98}) and Mehlert et al. (\cite{M+98}) find 
the ellipticals to be coeval in Coma and Fornax, and detect the presence 
of residual star formation in lenticulars only. 

\subsection{Subtracting the morphological segregation}

If the morphological segregation
results from the morphological segregation, $R_a$ should
also correlate with any other parameter dependent on the morphological
type. Such parameter could be the signature of a disk, as detected 
from image analysis (Scorza and Bender \cite{Sco95}) or from the 
rotational support (\rotsup, where $V_{\rm max}$ is the observed 
maximum rotational velocity).
Both of these parameters are affected by projection effects and are available
only for a restricted subsample. We will use \rotsup, taken from HYPERCAT.

In our sample, \rotsup is available for 272 galaxies: 157 ellipticals
and 115 lenticulars, the mean values are respectively: $<\rotsup> =0.27 \pm0.02$ and 
$0.43 \pm0.04$ (The corresponding mean \lrho are: $-0.46 \pm0.04$ and $-0.50 \pm0.04$). 
The difference is clear and shows that \rotsup is fairly 
correlated with the morphology, but the morphology-density relation is here very
marginal.

Because of the smaller statistics the \lrho -- $R_a$ correlation is much noisier
than in Fig.~\ref{fig1}, but the effect is still marginally observed in the 
subsample: The
mean \lrho for $R_a<0$ is $-0.50 \pm0.03$ (162 galaxies) and $-0.44
\pm0.04$ for $R_a\geq 0$
(114 galaxies). 
The corresponding mean \rotsup are: $0.328\pm0.023$ and $0.288\pm0.025$.
The difference is marginally significant $0.040\pm0.035$

The \rotsup -- \lrho correlation, represents the morphology -- density
relation, it is fitted as: $  \lrho \approx -0.25 \rotsup$. It predicts a
difference in \lrho between the two subsamples,  $R_a<0$ and $R_a>0$,
of: $0.010 \pm0.015$. This is about $1/4$ of the
observed effect. However, taking into account the large uncertainties,
it is not possible to reject the possibility that the totality of
the stellar population segregation
may be due to the morphology -- density relation.

Repeating the analysis on $R_f$ reproduces the dynamical effect found by
Prugniel \& Simien (\cite{PS94}) and does not help to understand the origin of
the population segregation.

\section{Conclusion}\label{sect_concl}

Analyzing the residuals to the \mg -- \sig and Fundamental Plane
relations as a function of the large scale density of the environment,
we find a segregation of the stellar population. The early-type galaxies
with a line strength index \mg weaker than expected from their value of \sig
are preferentially found in low-density regions. 
These galaxies are
likely to contain an excess of young stars, presumably formed after
a gravitational encounter or merging event that occurred in the last 
gigayears. This extends previous results to very sparse environments.

However, we cannot rule out that this effect is a by-product of the
morphological segregation.
Indeed, even our subsample of bona-fide ellipticals may still be contaminated 
by some lenticular galaxies seen almost face-on or with a weak disk, and the
observed segregation could be due to residual star formation in
these galaxies expected to be found mostly in low-density regions.
We tried to use the observed rotational support,
\rotsup, to parameterize the morphology, as it traces the presence of a disk.
The morphology-density relation seems to account for part 
of the segregation observed on stellar population, but the data are still 
too noisy to exclude that it is the totality. New data on a larger
sample are needed.

\begin{acknowledgements}

   We are grateful to the telescope operators at Observatoire de
   Haute-Provence for their help during the observations.
   VG thanks the CRAL-Observatoire Astronomique de Lyon for an
   invited-astronomer position.
We thank Guy Worthey for remarks that have improved the paper.

\end{acknowledgements}

\end{document}